\documentclass[letterpaper, 10 pt, conference]{ieeeconf}  

\IEEEoverridecommandlockouts                              

\usepackage{graphics} 
\usepackage{epsfig} 
\usepackage{mathptmx} 
\usepackage{times} 
\usepackage{amsmath} 
\usepackage{amssymb}  
\usepackage{color}

\title{\LARGE \bf
Automatic Preprocessing and Ensemble Learning \\ for Low Quality Cell Image Segmentation
}

\author{Sota Kato$^{1}$ and Kazuhiro Hotta$^{2}$
\thanks{$^{1}$The Department of Electrical and Electronic Engineering, Meijo University, 
        Nagoya-shi, 468–8502 Japan
        {\tt\small 150442030@ccalumni.meijo-u.ac.jp}}%
\thanks{$^{2}$The Department of Electrical and Electronic Engineering, Meijo University, 
        Nagoya-shi, 468–8502 Japan
        {\tt\small kazuhotta@meijo-u.ac.jp}}%
}

\begin{document}

\maketitle
\thispagestyle{empty}
\pagestyle{empty}

\begin{abstract}

We propose an automatic preprocessing and ensemble learning for segmentation of cell images with low quality.
It is difficult to capture cells with strong light. 
Therefore, the microscopic images of cells tend to have low image quality but these images are not good for semantic segmentation. 
Here we propose a method to translate an input image to the images that are easy to recognize by deep learning. 
The proposed method consists of two deep neural networks. 
The first network is the usual training for semantic segmentation, and penultimate feature maps of the first network are used as filters to translate an input image to the images that emphasize each class.
This is the automatic preprocessing and translated cell images are easily classified.
The input cell image with low quality is translated by the feature maps in the first network, and the translated images are fed into the second network for semantic segmentation. 
Since the outputs of the second network are multiple segmentation results, we conduct the weighted ensemble of those segmentation images.
Two networks are trained by end-to-end manner, and we do not need to prepare images with high quality for the translation. 
We confirmed that our proposed method can translate cell images with low quality to the images that are easy to segment, and segmentation accuracy has improved using the weighted ensemble learning.

\end{abstract}

\section{INTRODUCTION}
In recent years, segmentation task that assigns the class label to each pixel in an image is important in the field of medical and biological images [1-6]. 
In particular, cell image segmentation tends to be subjective because it has been done manually, but deep learning can get objective results and it received attracting attention. 
Many  segmentation  methods  have  been  proposed [7-9], but the segmentation accuracy of those methods depends on the quality of input images. 
Especially, in the case of cell images, cell images are low quality because cells die by a strong light.

To improve the segmentation accuracy, it is important to use preprocessing that deep learning can easily understand. 
However, there is little good preprocessing for deep learning.
In terms of clarifying the images, there is a method called super-resolution. 
Many super-resolution methods [10-17] using deep learning have been proposed. 
However, those super-resolution methods require teacher images. 
It has required a lot of time and computational cost to prepare teacher images.

In this paper, we propose an automatic preprocessing method for cell image segmentation by deep learning. 
Our proposed method consists of two deep neural networks. 
The first network is for semantic segmentation, and penultimate feature maps in the first network is used as filters to translate an input image to the images that are easy to segment. 
The number of channels of penultimate feature maps is the same as segmentation classes, and the input cell image is translated to multiple images that emphasizes each class.
The second network is for segmenting the images generated by the first network. 
The input cell image with low quality is translated by the filter, and the translated image is fed into the second network for segmentation. 
In addition, we propose automatic weighted ensemble learning to aggregate multiple segmentation images that the first network and second network generated.
By using automatic weighted ensemble learning, suitable weights are determined automatically and segmentation accuracy is further improved.
Fig.1 shows the example of automatic preprocessing and ensemble learning by the proposed method.

In experiments, we evaluated our method on two cell segmentation datasets which distinguishes cell images into multiple categories.
We compared our method with conventional methods with various preprocessing, and we confirmed that our proposed method could translate cell images with low quality to the images that were easy to segment, and the segmentation accuracy has improved.

This paper is organized as follows. 
Section 2 describes related works. 
Section 3 explains the details of the proposed method. 
We describe the datasets and evaluation method in section 4.
Section 5 shows the experimental results. 
Finally, we describe our summary and future works in section 6.

\begin{figure}[t]
    \centering
    \includegraphics[scale=0.25]{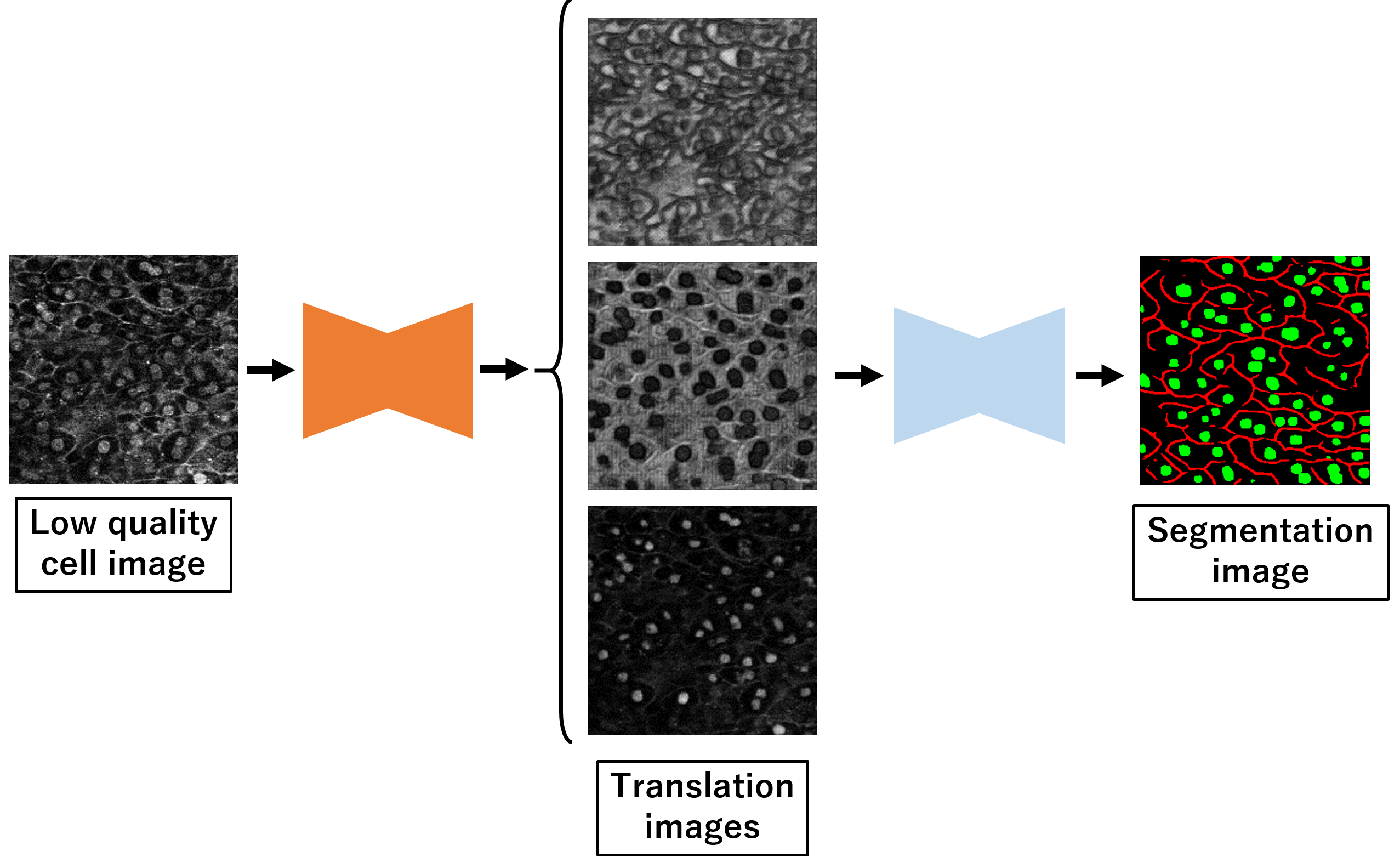}
    \caption{Example of automatic preprocessing and ensemble learning. The input cell image with low quality is translated to the images that are easy to segment through the first network, and aggregated by ensemble through the second network and a 3D convolution layer.}
    \label{fig:my_label1}
\end{figure}
\begin{figure*}[t]
    \centering
    \includegraphics[scale=0.45]{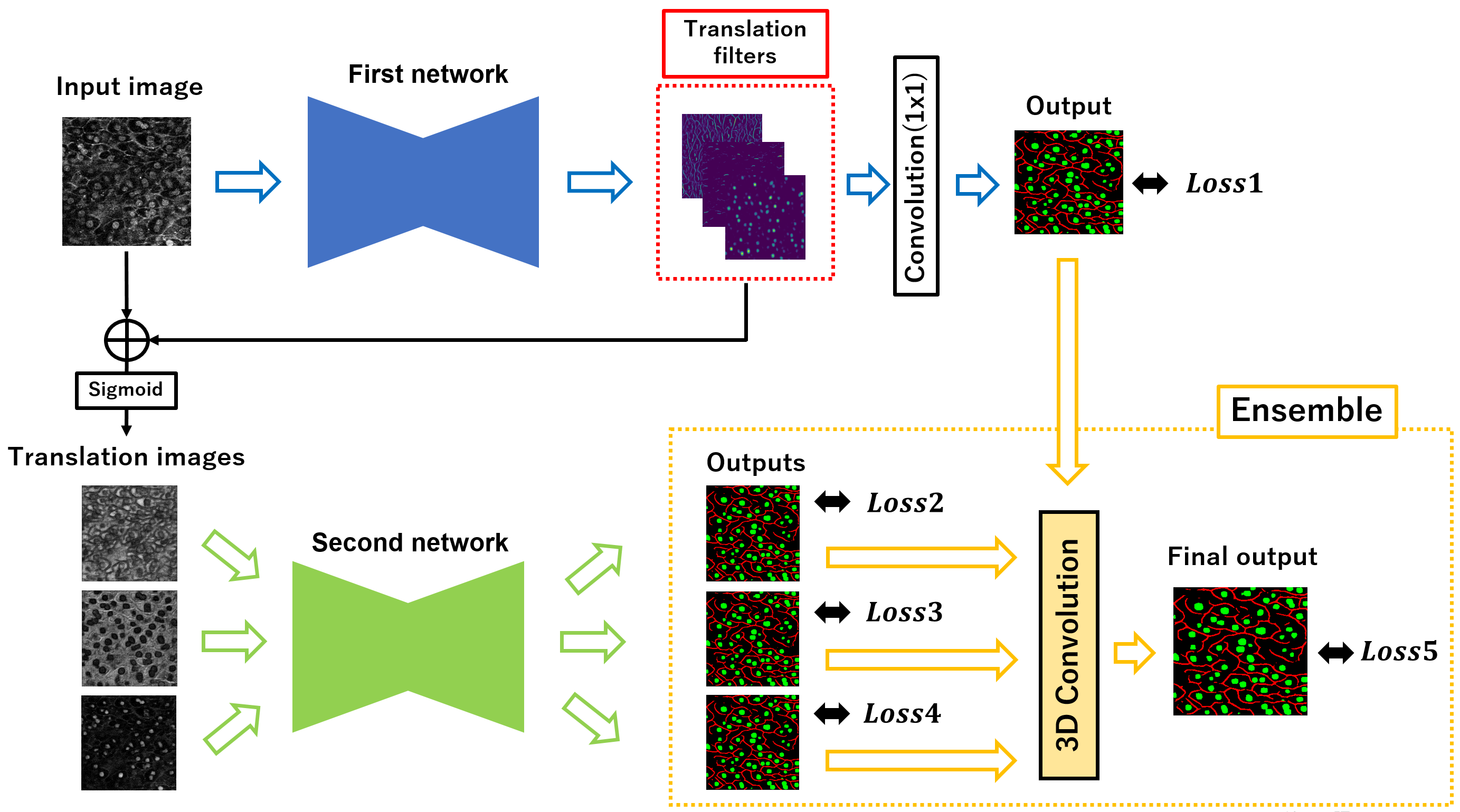}
    \caption{Overview of the proposed method using automatic preprocessing and the weighted ensemble learning. In segmentation of three classes, we use three translation filters. Each translation filter is added to an input cell image, and we obtain three translation images. Each translation image emphasized objects related to the segmentation result. Translated images are fed into the second network one by one for segmentation and we compute the loss for all segmentation images.}
    \label{fig:my_label1}
\end{figure*}

\section{Related works}
\subsection{Biologicacl image segmentation}

In cell biology, segmentation is a very important task because segmentation result is easy for us to understand.
In recent years, the segmentation accuracy has been improved by using deep learning. 
U-Net[18] is a famous segmentation method in cell biology and medical image processing. 
It is an encoder-decoder structured network. 
In encoder, the features of an input image are extracted by convolutions. 
Fine information such as correct position of objects is lost by down-sampling. 
In decoder, skip connection is introduced at each resolution. Skip connection is to concatenate the feature maps obtained by a encoder with the feature maps with the same resolution at a decoder. 
As a result, fine information and correct position which are lost in feature extraction process can be used effectively.

Although many segmentation methods based on U-Net for improving the accuracy have been proposed [19-21], the accuracy of them depends on the quality of input images. 
Thus, the preprocessing for segmentation with deep learning is required.

\subsection{Super resolution}

As similar concept to our proposed method, there are super-resolution methods using deep learning. 
Many methods have been proposed using deep convolutional neural network [10,11], Generative Adversarial Network (GAN) [12,13] and attention [14,15]. 
However, those methods require teacher images and it takes a lot of cost and time to prepare them. 
There is also a problem about GPU memory because the network must enlarge the size of image. 
Recently, unsupervised super-resolution methods have also been proposed [16,17] but the image quality is not sufficient. It is difficult to use it for preprocessing of microscope images with low quality. 

Therefore, we propose an automatic preprocessing method that CNN generates a filter for translating an input image to the image that can be easily segmented.

\section{Proposed method}

\begin{figure*}[t]
    \centering
    \includegraphics[scale=0.45]{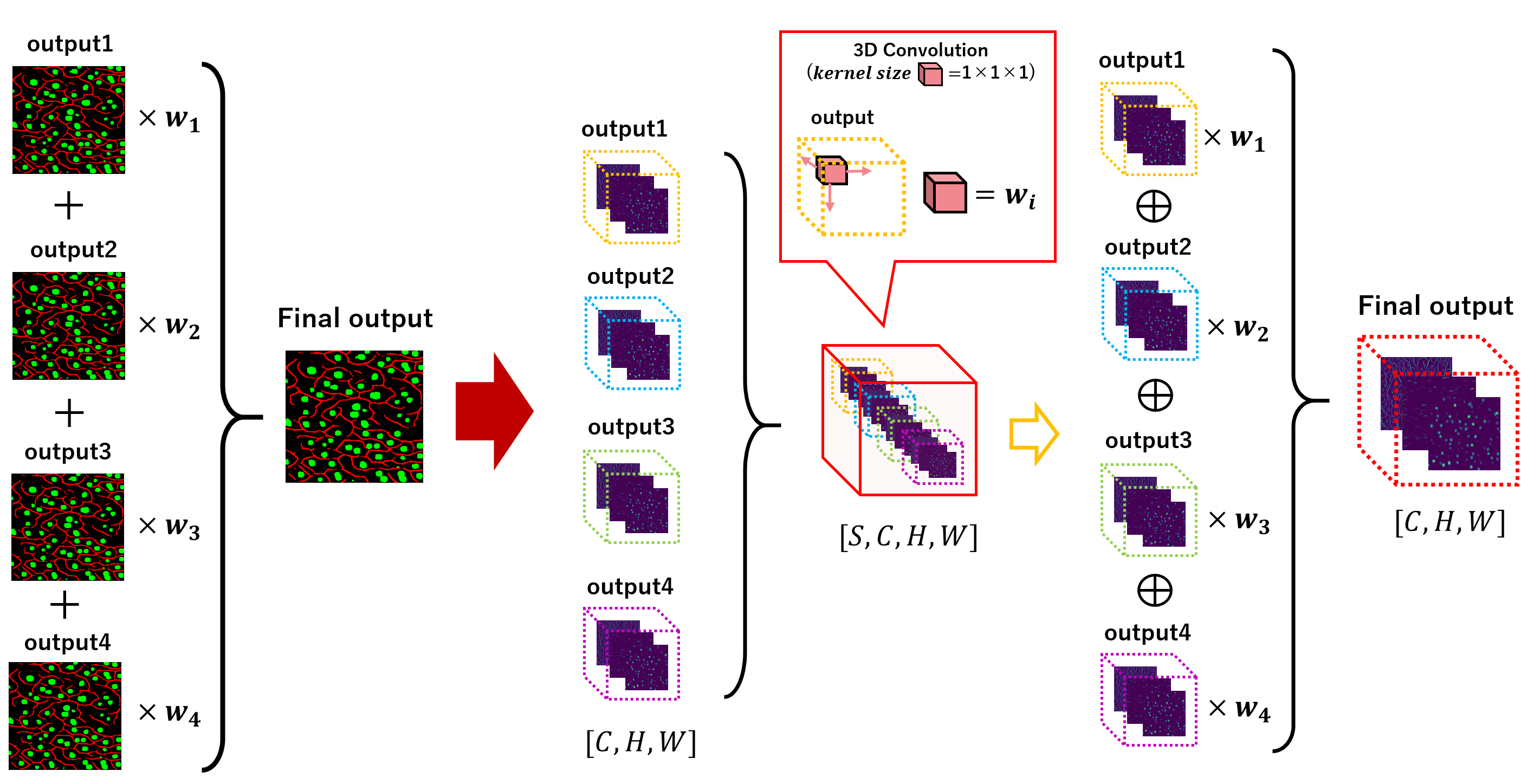}
    \caption{The architecture of weighted ensemble learning using 3D convolutional layer. In segmentation of three classes, we prepare four segmentation outputs. The first network generates one output and the second network generates three outputs. To aggregate outputs and generate one segmentation result, we use the weighted ensemble learning. Weights are determined automatically by 3D convolution when training networks.}
    \label{fig:my_label1}
\end{figure*}
Fig.2 shows the overview of the proposed method. 
To make the input image more suitable for segmentation, we propose an unsupervised image translation method using deep learning. 
First, filters for translating an input image to suitable image for segmentation are generated by penultimate feature maps in the first network.
The size and number of channels in the generated filters are the same as the input image and the number of segmentation classes.
Since the first network outputs the segmentation image, the generated filters have information that are useful for segmentation and emphasize objects related to the segmentation result.
This is automatic preprocessing using deep learning.
We do not require ground truth with high quality for generating filters.
How to generate filters for an input image to improve the segmentation accuracy is also trained automatically. 
There is no negative information in the filters because we use ReLU function before the filter outputs. 
Therefore, we see that the filters corrected the luminance in the input image. 
Since the luminance value is too large to interfere with learning, it is normalized from 0 to 1 by using a sigmoid function. 

The generated filters are added to the input image, and translated images are fed into the second network for segmentation and the loss for segmentation with the ground truth image is minimized. 
Since the number of translated images is the same as the segmentation classes and we feed each translated image to the second network independently, the second network outputs multiple segmentation images.
Segmentation results obtained from each translated image are different because of each translated image is different from the original input image.
Finally, segmentation images generated by both of the first and second networks are aggregated using the weighted ensemble learning.
We reduce the total error by aggregating some segmentation outputs. 
Both networks for filter generation and segmentation are trained simultaneously to generate segmentation result with high accuracy. 
We use cross entropy loss for all outputs.
Equation (1) shows the final loss where \textit{Loss$_1$} is the error of the first network, \textit{Loss$_2$, Loss$_3$, Loss$_4$} are errors of the second network and Loss$_5$ is the error of aggregated outputs by ensemble learning.
\begin{eqnarray}
  Loss  =  Loss_1 + Loss_2 + Loss_3 + Loss_4 + Loss_5
\end{eqnarray}

\subsection{Weighted ensemble learning using 3D convolution}

The aim of ensemble learning is that multiple segmentation images generated by the first and second networks are aggregated to one segmentation result to improve the segmentation accuracy.
There are two kinds of ensemble learning; normal average and weighted average.
Generically, the weighted average is better than normal average because it is possible to assign large weight to important elements.
However, it is difficult to determine suitable weight values.
Therefore, we propose the weighted ensemble learning which determines weights automatically using 3D convolutional layer.

Fig.3 shows the architecture of the weighted ensemble learning. 
The size of each segmentation result of the first and second network is $[C \times H \times W]$ where $H$ and $W$ are the height and width of a output image and $C$ is the number of classes. 
All outputs are aggregate as $[S \times C \times H \times W]$ where $S$ is the number of outputs. 
Here, we use a 3D convolutional layer with 1×1×1 kernels, 1 stride and 0 padding.
This is called point-wise convolution.
The point-wise 3D convolution calculates only the channel direction. 
Since aggregated array $[S \times C \times H \times W]$ is four dimensions, it is possible to adopt point-wise 3D convolution to the aggregated array by replacing $[S]$ in the aggregated array to a channel direction.
Therefore, we can assign a weight $w_i$ in Fig.3 to each segmentation output $[C \times H \times W]$ through training, and we are able to generate the final segmentation result from $[S]$ results automatically.

\begin{figure}[t]
    \centering
    \includegraphics[scale=0.28]{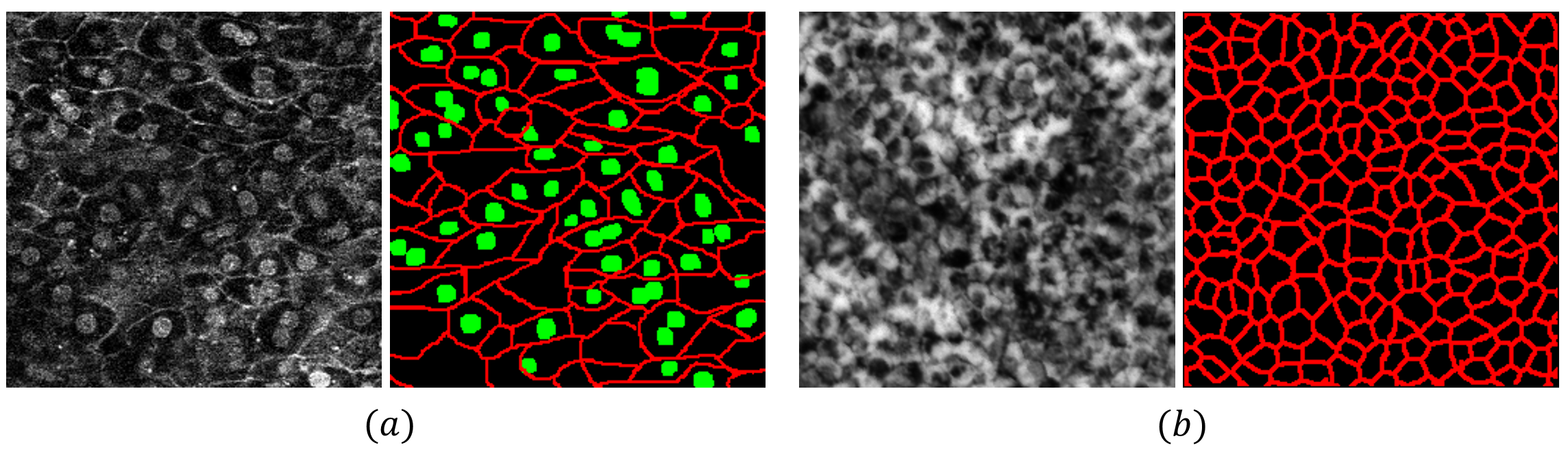}
    \caption{Examples of cell images their ground truths in two datasets. $(a)$ is the cell image of mouse liver with three class labels; cell nucleus (green), cell membrane (red) and cytoplasm (black). $(b)$ is the human iRPE cell image which includes cell membrane (red) and background (black).}
    \label{fig:my_label1}
\end{figure}
\section{Overview of experiments}
\subsection{Dataset}
\begin{table*}[t]
    \centering
    \caption{Comparison between conventional method and our proposed method on cell image dataset of mouse liver}
    \scalebox{0.90}{
    \begin{tabular}{c||c|ccc||c|ccc} \hline
    Evaluation & \multicolumn{4}{c||}{IoU} & \multicolumn{4}{c}{Dice} \\ 
    \hline
    Methods & {Mean} & {cytoplasm} & {membrane} & {nucleus} & {Mean} & {cytoplasm} & {membrane} & {nucleus}\\ 
    \hline
        U-Net [18]&57.90(±1.33)&71.38(±4.80)&40.32(±3.84)& 62.01(±2.65)&72.36(±0.73)&83.21(±3.31)&57.36(±3.95)& 76.52(±2.01)\\
        U-Net++ [19]&57.29(±1.23)&71.38(±4.67)&39.80(±4.20)& 60.68(±3.16)&71.84(±0.59)&83.21(±3.22)&56.82(±4.35)& 75.48(±2.42)\\
        Attention U-Net [25]&58.06(±1.21)&\textbf{71.98(±3.83)}&40.00(±3.49)& 62.19(±2.79)&72.46(±0.69)&\textbf{83.65(±2.58)}&57.06(±3.64)& 76.65(±2.08)\\
        Proposed method&\textbf{59.53(±1.72)}&71.91(±3.91)&\textbf{43.27(±2.32)}& \textbf{63.42(±3.38)}&\textbf{73.84(±1.06)}&83.60(±2.64)&\textbf{60.36(±2.26)}& \textbf{77.56(±2.50)} \\
    \hline
    \end{tabular}
    }
\end{table*}
\begin{table}[t]
    \centering
    \caption{Ablation studies for ensemble learning}
    \scalebox{0.80}{
    \begin{tabular}{c||c|ccc} \hline
    Evaluation & \multicolumn{4}{c}{IoU} \\ 
    \hline
    Methods & {Mean} & {cytoplasm} & {membrane} & {nucleus}\\ 
    \hline
        U-Net [18]&57.90(±1.33)&71.38(±4.80)&40.32(±3.84)& 62.01(±2.65)\\
        w/o Ensemble&59.27(±1.81)&\textbf{72.40(±3.92)}&42.44(±2.27)& 62.98(±3.57)\\
        Ensemble(fixed)&59.41(±0.73)&72.01(±0.73)&42.94(±0.73)& 63.26(±0.73)\\
        Ensemble(automated)&\textbf{59.53(±1.72)}&71.91(±3.91)&\textbf{43.27(±2.32)}& \textbf{63.42(±3.38)}\\
    \hline
    \end{tabular}
    }
\end{table}
\begin{table*}[t]
    \centering
    \caption{Comparison between conventional method and our proposed method on human iRPE cell images}
    \scalebox{1.10}{
    \begin{tabular}{c||c|cc||c|cc} \hline
    Evaluation & \multicolumn{3}{c||}{IoU} & \multicolumn{3}{c}{Dice} \\ 
    \hline
    Methods & {Mean} & {background} & {membrane} & {Mean} & {background} & {membrane} \\ 
    \hline
        U-Net [18]&62.86(±0.37)&\textbf{76.09(±0.09)}&49.64(±0.69)&76.38(±0.32)&\textbf{86.42(±0.06)}&66.34(±0.61)\\
        U-Net++ [19]&62.33(±0.68)&74.97(±0.68)&49.69(±0.89)&76.04(±0.54)&85.69(±0.45)&66.38(±0.79)\\
        Attention U-Net [25]&63.95(±0.45)&76.07(±0.40)&51.82(±0.53)&77.34(±0.35)&86.41(±0.26)&68.26(±0.46)\\
        Proposed method&\textbf{64.05(±0.45)}&75.92(±0.33)&\textbf{52.19(±0.75)}& \textbf{77.45(±0.37)}&86.31(±0.22)& \textbf{68.58(±0.65)} \\
    \hline
    \end{tabular}
    }
\end{table*}

We use 50 cell images of mouth with ground truth attached by Kyoto University [22]. 
The ground truth image includes 3 labels; cytoplasm, nucleus and membrane.
Images and ground truth sizes are 256×256 pixels. 
35 images are used for training, 5 images are used for validation and the remaining 10 images are for test. We use 5-fold cross validation.

We also evaluate our method on the other datasets. 
We use absorbance microscopy images of human iRPE cells [23]. 
The ground truth image includes 2 kinds of labels; background and membrane.
The images are split into 1,032 regions of 256×256 pixels and the corresponding ground-truth.
688 images are used for training, 44 images are used for validation and the remaining 300 images are for test. We used 3-fold cross validation.

Examples of cell images in two datasets and their ground truths are shown in Fig.4. 
Left two columns shows cell image of mouse liver and its ground truth includes three class labels; cell nucleus (green), cell membrane (red) and cytoplasm (black). 
Right two columns shows the human iRPE cell image dataset which includes cell membrane (red) and background (black).

\subsection{Evaluation}

The segmentation accuracy of each class is evaluated by Interactive over Union (IoU). 
IoU computes the overlapping ratio between the predicted result and ground truth. 
Since the number of pixels in each class is different, we use mean IoU (mIoU) as final evaluation measure.

In experiments, we evaluate the proposed method and segmentation network without preprocessing network in order to show the effectiveness of our proposed unsupervised automatic preprocessing. 
Original U-Net [18] is used as two segmentation networks because it is often used for segmentation of cell images [20].
The batch size for training is set to 16 and Adam \textit{(betas = 0.9, 0.999)} is used as the optimization. The learning rate is set to $1 \times 10 ^ {-3}$. 
We train all networks till 300 epochs.
The images are normalized between 0 and 1 and no other preprocessing is done.
All experiments are done under exactly same conditions of dataset size, optimizer and the number of epochs.

\section{Experiment results}
\begin{figure*}[t]
    \centering
    \includegraphics[scale=0.38]{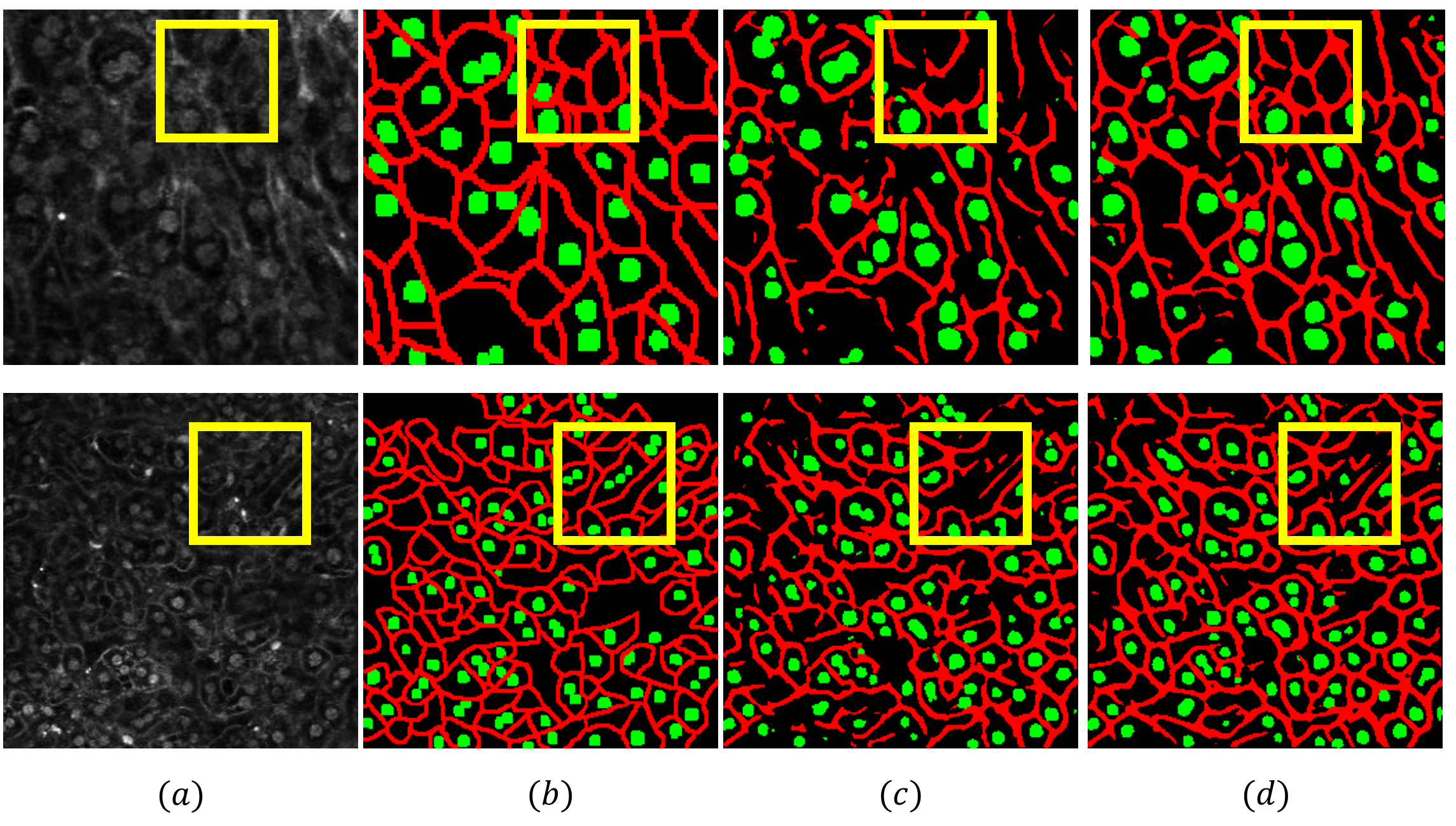}
    \caption{Comparison results on cell image dataset of the mouse liver. (a) Input cell image. (b) Ground truth. (c) Conventional method (U-Net). (d) Proposed method.}
    \label{fig:my_label1}
\end{figure*}
\begin{figure*}[t]
    \centering
    \includegraphics[scale=0.45]{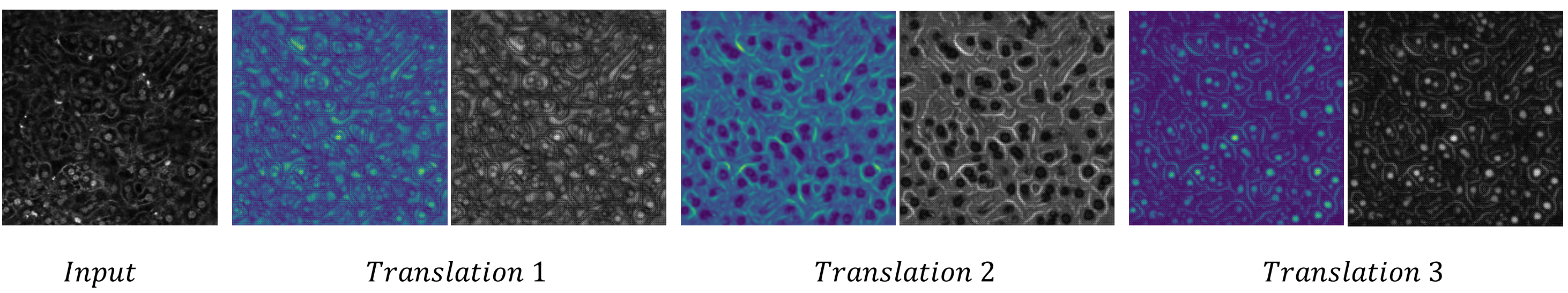}
    \caption{Translation filters automatically generated and the results of preprocessing for segmentation. Left column are the translation filters and right column are the preprocessing images.}
    \label{fig:my_label1}
\end{figure*}
\subsection{Results on cell images of the mouse liver}

Table 1 shows the results of segmentation for cell images of the mouse liver. 
We evaluated conventional methods [18,19,25] and the proposed method with the ensemble learning automated weights (automated). 
The proposed method with the ensemble learning automated weights improved IoU about 1.41\% for the cell nucleus and 2.95 \% for the cell membrane in comparison with U-Net without preprocessing.
Dice coefficient of our method has improved about 1.04\% for the cell nucleus and 3.00 \% for the cell membrane.
Totally, mIoU has improved about 1.63\% and mDice has improved about 1.48\%. 
Surprisingly, the ground truth was not used for translated images but adequate preprocessing for segmentation was realized. 
This result demonstrated the effectiveness of the proposed automatic preprocessing.

Table 2 shows the results of ablation studies for ensemble learning.
We evaluated the proposed method without the ensemble learning, with the ensemble learning using the fixed weights and that using automated weights.
Without the ensemble learning, the output of the second network is only one segmentation image for the final result.
Fixed weights were defined as $w_i=1$.
The proposed method without the ensemble learning improved mIoU about 1.37\% in comparison with U-Net without preprocessing.
This result demonstrated the effectiveness of transformed images.
Furthermore, the proposed method with the ensemble learning using automated weights improved mIoU about 0.12\% in comparison with the ensemble learning fixed weights.
The ensemble learning determined weights automatically was more effective than fixed weight ensemble learning.

Fig.5 shows the segmentation results. 
When we focused on yellow squares in Fig.5, the proposed method could segment cell membrane that conventional U-Net could not segment well. 
Fig.6 shows the translation filters automatically generated (left column) and the results of preprocessing for segmentation (right column). 
In translation filters, yellow means large weight and blue means small weight. 
These images show that the necessary information for segmentation in the image has emphasized and the unnecessary information has suppressed.
These results demonstrated the effectiveness of our translation filter generated automatically.
In images before translation, it was not possible to confirm cell nucleus and membrane with small brightness. 
However, in images after preprocessing, cell nucleus with small brightness became clearly, and cell membrane which has become like the noise is more clearly identified. 
In particular, the cell membrane existed in the noisy part was difficult for the human to classify. 
However, we confirmed that the cell membrane was more emphasized by the filter and the generated filter was suitable for segmentation.
The improvement IoU on cell membrane was 2.62\%.

\subsection{Results on human iRPE cell images}

We also evaluated our method on the other cell membrane datasets. Table 3 shows the results of the U-Net without preprocessing and the proposed methods. 
The proposed method with the weighted ensemble learning improved IoU about 2.55 \% and Dice coefficient was improved about 2.24 \% for the cell membrane. Totally, mIoU was improved about 1.19\% and mDice was improved about 1.07\% in comparison with the U-Net without preprocessing. Experimental results demonstrated the proposed method was effectiveness for the other cell image dataset.

Fig.7 shows the segmentation results on the human iRPE cell images. 
Figures show that our method worked well even if input images were much different from the previous experiment, and the segmentation accuracy by the proposed method is better than the U-Net without preprocessing. 

\section{Conclusion}

We proposed the segmentation method using automatic preprocessing and ensemble learning. 
By the experiments using actual cell images, we translated input images to the images that are easy to segment and mIoU was improved about 1.63\% in comparison with segmentation network without preprocessing. 
We also demonstrated that our method worked well on the other cell image dataset. 
Since we learned the first network using ground truth for segmentation, translation filters were affected by ground truth for segmentation.
However, by using unsupervised segmentation methods [24], it may generate a filter which is not affected segmentation teacher and has diversity. 
This is a subject for future works.

\section*{ACKNOWLEDGMENT}

This work was partially supported by KIOXIA Corporation.


\end{document}